\documentclass[
    ,final            
  ]
  {aipproc}

\layoutstyle{6x9}

\begin{document}

\title{Supernova Asymmetries}

\classification{\texttt{http://www.aip..org/pacs/index.html}}

\author{J. Craig Wheeler}{
  address={Department of Astronomy, University of Texas at Austin}
}

\author{Justyn R. Maund}{
  address={Department of Astronomy, University of Texas at Austin}
}

\author{Shizuka Akiyama}{
  address={Lawrence Livermore National Laboratory}
}

\begin{abstract}

All core collapse supernovae are strongly aspherical. The ``Bochum 
event," with velocity components displaced symmetrically about the 
principal H$\alpha$ line, strongly suggests that SN 1987A was a 
bi-polar rather than a uni-polar explosion. While there is a 
general tendency to display a single prominant axis in images
and spectropolarimetry, there is also growing evidence for 
frequent departures from axisymmetry. There are various mechanisms
that might contribute to large scale departures from spherical
symmetry: jet-induced processes, the spherical shock accretion 
instability (SASI) and associated phenomena, and non-axisymmetric 
instabilities (NAXI). The MRI gives inevitable production of large 
toroidal magnetic fields. In sum: no $\Omega$ without B. The role
of magnetic fields, non-axisymmetric instabilities, and of the 
de-leptonization phase are discussed. 

\end{abstract}

\maketitle


\section{Introduction}

All core collapse events with adequate spectropolarimetric 
observations have proven to be polarized and hence to depart
from spherical symmetry in some substantial way (Wang et al.
2001, 2003; Leonard et al. 2001a,b, 2006). Much of the spectropolarimetry
shows a tendency for the data to be distributed along a single
locus in the plane defined by the Stokes parameters Q and U. 
We are coming to understand, however, that departures from a
single locus are rather common, and possibly systematic. This
implies a breakdown in axisymmetry that must be understood. 
Although this is becoming generally recognized with recent 
detailed spectropolarimetric studies of distant supernovae,
SN 1987A provided the first evidence (Cropper et al. 1988; Jeffery 1991).

On the theoretical side, core collapse generically produces a 
structure in the proto-neutron star that has a strongly negative 
angular velocity gradient and hence is unstable to the magnetorotational 
instability (Velikov 1959; Chandrasekhar 1960; Acheson \& Gibson 1978; 
Balbus \& Hawley 1991, 1998). The MRI will exponentially grow the 
magnetic field on the rotational timescale by a self-induced turbulent 
dynamo process and produce strong, primarily toroidal magnetic fields 
in the proto-neutron star (Akiyama et al. 2003). It is not truely 
self-consistent to consider rotating core collapse without the 
concomitant growth, saturation, and subsequent evolution of this 
magnetic field. The ultimate problem is complex, involving rotation, 
magnetic fields, and neutrino transport, but it involves very 
interesting, and still underexplored, physics.

\section{Spectropolarimetry}

The first supernova for which good photometric and spectropolarimetric
data were obtained was SN 1987A. This data has still not been
adequately explored and we can view it now in the context of
the growing data base of more distant, but ever better studied
supernovae. 
Jeffery (1991) summarized the photometric polarimetry obtained
on SN 1987A (Fig. 1). Both B band and V band data showed a slow growth
to a polarization of 0.4 - 0.7\% by day 30 - 40. The polarization
then declined to a value near 0.2 - 0.3\% by day 100. Around
day 110, when the major maximum gave way to the exponential 
radioactive tail, the polarization jumped to 1.3 to 1.5\% and
then slowly dropped back to around 0.2 to 0.4\% by day 200. 
This jump is clearly associated with the photosphere receding
through the outer hydrogen envelope and revealing the inner
core. This behavior was caught again for the Type IIP SN 2005dj
by Leonard et al. (2006). SN 1987A gave clear evidence that the 
inner machine of the explosion was strongly asymmetric, evidence
that has proven ubiquitous with current, systematic observations.
Another remarkable fact is that the polarization angle did not
waver through this whole evolution, including the large spike
in polarization.  SN 1987A ``pointed" in a certain direction and 
maintained that orientation througout its development (Wang et al. 
2002). This cannot be due to Rayleigh-Tayler nor Richtmyer-Meshkov 
instability.  Other, large scale, systematic, directed asymmetries 
must be at work. The ``Bochum event," with velocity components 
displaced symmetrically about the principle H$\alpha$ line, strongly
suggests that SN 1987A was a bi-polar explosion (Hanuschik et al. 
1989; Wang et al. 2002).


\begin{figure}
  \includegraphics[height=.3\textheight]{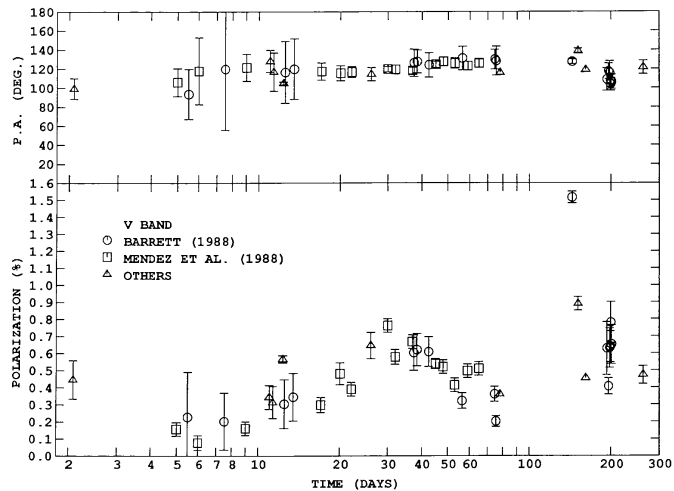}
  \caption{Evolution of the V-band polarization, 
  and associated polarization angle, of SN 1987A from Jeffery (1991).}
\end{figure}

On the other hand, the excellent spectropolarimetry of Cropper et al.
(1988; Fig. 2) showed that as data is tracked as a function of wavelength
over spectral features, the polarization angle does sometimes
change with wavelength, giving rise to ``loops" in the plane of
the Stokes parameters, Q and U. This means that there must be
some substantial departure from axisymmetry imposed on the
overall ``pointed" behavior revealed by the photometric polarimetry.
The loops are a locus with respect to wavelength, which itself
is a probe of velocity slices in the homologously expanding matter.
This polarimetric behavior thus gives a rich phenomenology that is 
ripe in SN 1987A and other events for progress in physical understanding. 
These loops will give greater insight into the composition-dependent
three-dimensional structure of the ejecta. 


\begin{figure}
  \includegraphics[height=.3\textheight]{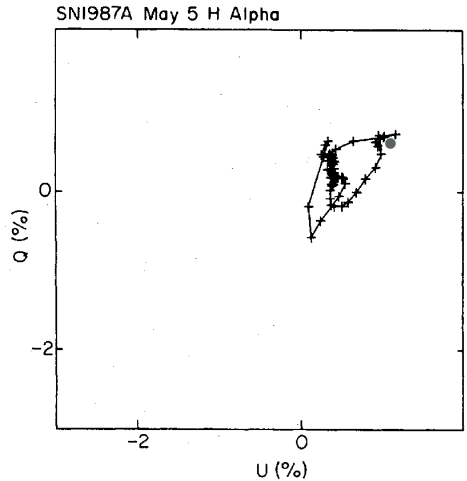}
  \caption{H$\alpha$, on the $Q-U$ plane, of SN 1987A 
  on May 5 1987 from Cropper et al. (1988).  The filled grey circle 
  is the ISP, effectively the origin of polarization intrinsic to 
  SN 1987A, on the $Q-U$ plane from Mendez (private communication).}
\end{figure}
   
Two other examples of non-axisymmetric loop structures in polarization
data are given in Maund et al. (2007a, b). Maund et al. (2007a)
discuss data on the Type IIb event SN 2001ig. Four days after
discovery, when the supernova was still in the H-rich phase, the
blended H$\alpha$/He I 6678 P-Cygni feature shows a distinct
loop in the Q/U plane, again signifying a systematic departure 
from axisymmetry (Fig. 3; left panel). In this case, the blending 
of the two lines plays a special role. Maund et al. (2007b) present 
data on the weird Type Ib/c SN 2005bf that resembled a helium-poor
Type Ic in early data, but developed distinct helium-rich
Type Ib features later (Wang \& Baade 2005; Folatelli et al. 2006).
Our observations on May 1, 2005, 34 days after the explosion,
18 days after the first peak in the light curve, and 6 days before
the second peak, show a distinct loop in the He I 5876 line
(Fig. 3; right panel). Related complex structures were revealed 
by the high-velocity Type Ic SN 2002ap (Wang et al. 2003). Thus 
although the sample is still small, evidence for non-axisymmetry 
may be ubiquitous.


\begin{figure}
\rotatebox{-90}{
  \includegraphics[width=7cm]{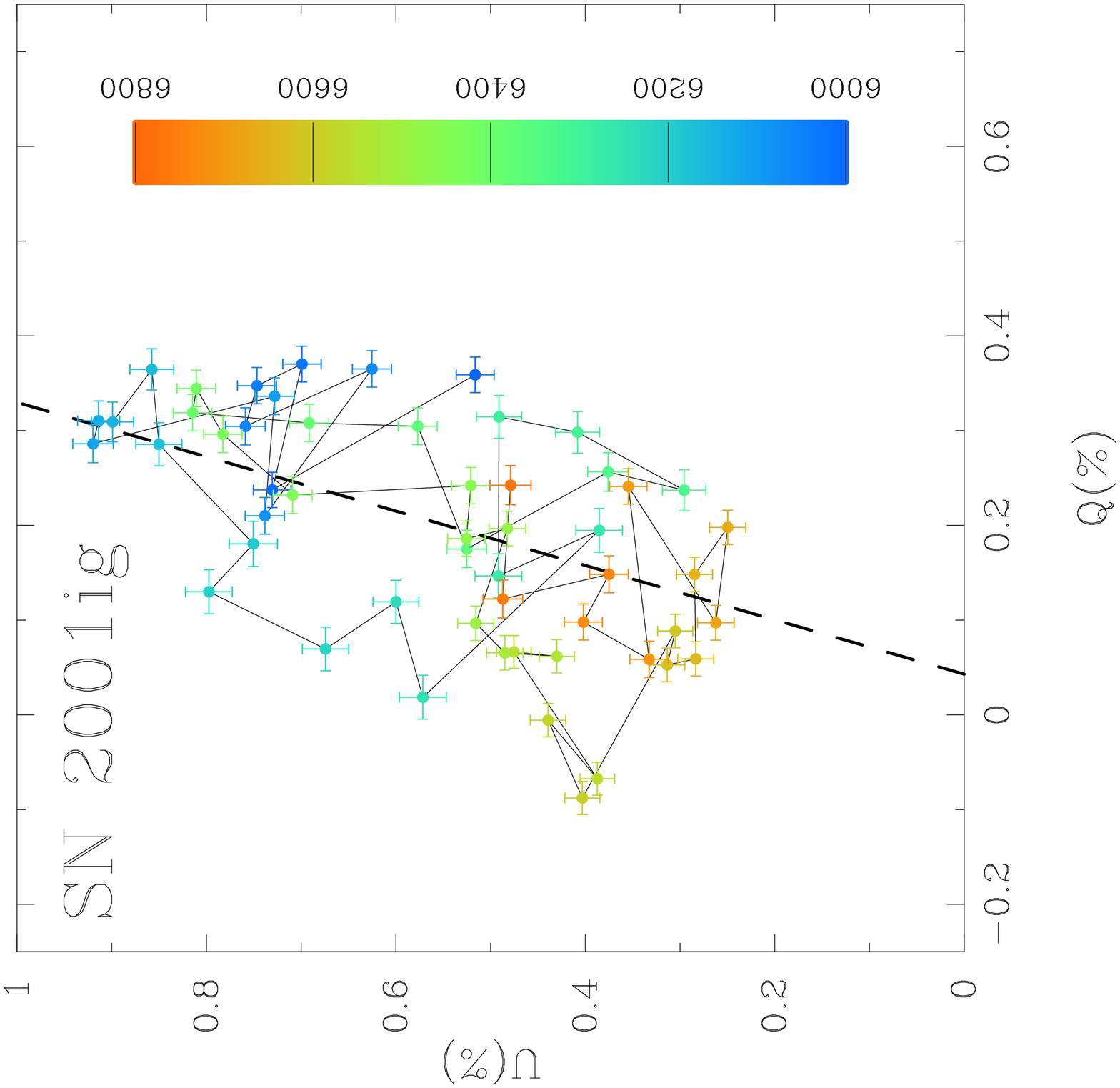}}
  \rotatebox{-90}{
  \includegraphics[width=7.1cm]{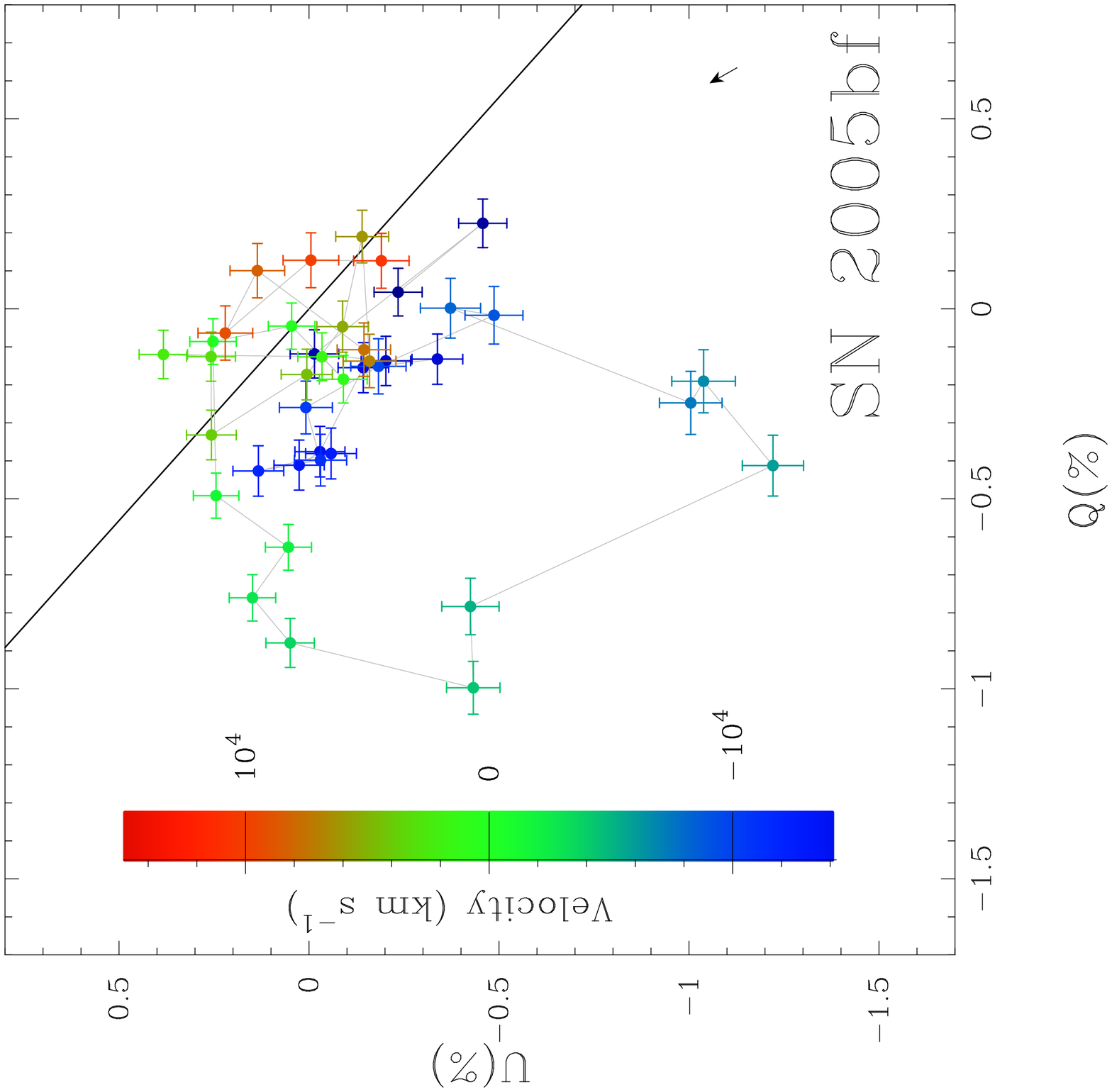}}
  \caption{$Q-U$ plane loops, corrected for the relevant 
  ISP, of He I 6678\AA/H$\alpha$\ in the Type IIb SN 2001ig and 
  He I 5876\AA\ in the Type Ib/c SN 2005bf  (Maund et al. 2007a; 
  Maund et al. 2007b).}
\end{figure}

A full understanding of the spectropolarimetry requires allowance 
for the background polarization of the interstellar medium
of our Galaxy, the host galaxy and, perhaps, the circumstellar 
environment of the supernova. Cropper et al. (1988) presented their 
data with no reference to this background. Jeffery (1991) quotes
Mendez (private communication) as determining the ISP to be
about Q = +0.36\% and U = +0.89\%. This component ``stayed constant
for about 6 months and seemed Serkowski-like." Care must be taken
since any ground-based point-spread function would include data from
all the background stars, some of which are Be stars that could
be intrinsically, and variably polarized themselves, a problem
for all extragalactic data. Taking the Mendez results at face value
gives a somewhat different perspective on the Cropper et al. data. 
For instance, as originally plotted, the Cropper et al. data give 
the impression that on June 3, after the transition to the radioactive 
tail and during the spike of maximum photometric polarization, the H$\alpha$
and Ca II IR features appear to show large ``loops" that are co-aligned, 
but in some sense mirror images of one another with H$\alpha$ 
extending to positive Q and U and Ca II extending to negative Q and U.
With the Mendez ISP point (Fig. 2), the June 3 H$\alpha$ loop is 
roughly symmetric in its major extent around the ISP point, but 
Ca II only extends to negative values. Both ``loops" appear to have 
a net displacement to the positive Q side of a line parallel to the 
major axis of the loop and passing through the ISP point. The 
similarities and differences between these polarization structures 
are worth study in considerable more depth.

Departure from axisymmetry and specifically the formation of
``loops" has been discussed by Kasen et al. (2003) in the
context of Type Ia supernovae and by Hoffman (2007) in
the context of Type IIn supernovae. These structures could
be caused by clumps of high-opacity material partially
obscuring a polarized photosphere or by coherent, non-axisymmetric,
composition-dependent, structures in the ejecta.  
 
Core collapse supernove, including SN 1987A, thus show strong
evidence for axisymmetry, but also significant departures from
axisymmetry. The issue thus arises as to what physical
processes drive the breakdown in spherical symmetry.

\section{The Physics of Asymmetry}

\subsection{Spherical Accretion Shock Instability: SASI}

Blondin, Mezzacappa \& DeMarino (2003) showed that the standing
shock that forms after core collapse will be unstable. There is
still controversy as to whether this is a purely 
acoustically-driven instability that propagates primarily azimuthally 
(Blondin \& Mezzacappa 2006) or whether there is some radially-coupled
vortical/acoustic feedback loop (Foglizzo et al. 2007). Burrows
et al. (2006) have argued that the SASI may lead to, but be
different from, a large scale instability in which g-modes 
are excited on the proto-neutron star surface that drive a 
large acoustic flux that could play a role in the explosion,
perhaps making a substantially axisymmetric, but uni-polar
explosion. As remarked above, SN 1987A, at least, appears to
have been truely bi-polar. There is, in any case, no doubt that the 
basic SASI instability exists. It must be taken into account in the 
physics of core collapse, and it will tend to introduce substantial
asymmetries.

Most of the simulations to date are 2D, non-rotating, and have no
magnetic fields. Some omit neutrino transport. An important step
was taken by Blondin \& Mezzacappa (2007) to investigate the SASI
in 3D. Remarkably, they found that a model with an initially
non-rotating iron core, and hence no net angular momentum, 
produced a proto-neutron star rotating at with a 50 msec period.
This comes about because the SASI in 3D produces substantial
lateral flows in the outer portions. To maintain conservation
of angular momentum, an equivalent, but oppositely directed,
angular momentum must be accreted onto the core. Young \& Fryer
(2007) find a similar behavior in 3D SPH calculations. This
sort of behavior is extremely interesting since it relaxes
the coupling between the rotational state of the original
iron core and the spin rate of the neutron star that
subsequently forms (see also Akiyama \& Wheeler 2005).  
  
\subsection{Non-Axisymmetric Instabilities (NAXI) in the Neutron Star}

Rotation with ratio of rotational to binding energy
T/|W| > 0.01 (P $\sim$ 20 - 30 msec) may be subject to non-axisymmetric 
instability (Andersson 1998; Shibata et al. 2003; Ott et al. 2005).
The thresholds, growth rates, and saturation, depend on the degree 
of differential rotation (Tohline \& Hachisu 1990; Rampp et al. 1998; 
Centrella et al. 2001; Imamura \& Durisen 2004; Ou et al. 2004) 
and will be affected by magnetic fields (Rezzolla, Lamb \& Shapiro 2000, 
2001a,b). A dynamic instability to a bar-like mode can occur for T/|W| 
as low as $\sim$ 0.2. The criterion T/|W| > 0.27 is neither necessary 
nor sufficient for dynamical instability (Shibata \& Sekiguchi 2005). 

Most work on non-axisymmetric instabilities considers the neutron 
star to be in complete isolation and ignores the magnetic field; 
both are potentially significant. In the current context, it is 
important to realize that for sufficiently rapidly rotating 
proto-neutron stars, these NAXI can occur deep inside the supernova 
material, before it has time to expand and dissipate. The MRI will 
grow magnetic field faster than all but dynamical bar-modes. This
means that the non-axisymmetric instabilities will usually occur in 
a magnetized medium. 

Ou \& Tohline (2006) examined the NAXI due to the presence of a 
co-rotation point within a ``resonant cavity" formed by an 
extremum in vorticity density in a differentially rotating 
structure that serves to reflect acoustic waves. Note that
while one-armed m = 1 spiral modes can be generated by this
mechanism and by the SASI, the azimuthal acoustic modes due 
to the SASI are not the same as those of the NAXI in a 
differentially-rotating proto-neutron star. 

It is important to note that the density distribution near 
the region of  peak shear and magnetic field in a proto-neutron 
star (the boundary of the homologous core) is approximately 
independent of whether a supernova has been successful or not in 
the interval 100 msec to 1 s after bounce. This means that the NAXI 
will interact with surrounding, magnetized matter. The result is 
likely to be the generation of significant magnetoacoustic flux that will
propagate out into the ambient matter, whether or not an explosion
has been initiated by some other mechanism (Wheeler \& Akiyama
2007). Note also that according to the results of Blondin
\& Mezzacappa (2007) and of Young \& Fryer (2007), non-rotating 
iron cores can potentially generate rotating neutron stars subject 
to NAXI, and hence to magnetoacoustic flux.  Wheeler
\& Akiyama (2007) estimate that for parameters appropriate to the
proto-neutron star, a luminosity $L_{mhd} \sim 10^{50 - 52}$ erg s$^{-1}$ 
could be radiated in magnetoacoustic flux on a time scale 
somewhat shorter than the de-leptonization time. At the larger end
of this range, corresponding to T/|W| near the upper limit of 0.14, 
the luminosity is quite competitive with neutrino heating rates. 

\subsection{The De-leptonization Phase}

The proto-neutron star formed at core bounce will radiate 
binding energy in neutrinos, contract, and spin up. The
time for de-leptonization is $\sim 10$ s to radiate most of
the neutrino energy. The time is shorter for the contraction
of the radius, the key factor in spinning up the neutron star.
The former is comparable to the time for the blast wave to propagate 
out of the C/O or He core.  This means that de-leptonization also 
does not occur in the vacuum of space, but takes place within 
the matter-filled environment in the center of the supernova, 
whether it is in the process of exploding or not.

The tendency for the de-leptonizing PNS to spin up will render it 
broadly susceptible to non-axisymmetric instabilities, the 
production of magnetoacoustic luminosity, and dissipation of rotation.
The evolution depends on the strength of that dissipation, but
in general the time for dissipation of the rotational energy 
by magnetoacoustic flux will be less than the time for 
de-leptonization, $\sim$ 10 s. Constraints on the energetics of
the de-leptonization phase are given by Wheeler \& Akiyama (2007).

\subsection{Jet-Induced Models}

The tendency for core collapse supernovae to follow a fixed
axis has motivated interest in jet-induced explosions (Khokhlov
et al. 1999). Rotation and magnetic fields are important
ingredients in core collapse and those factors commonly
lead to jets in many astrophysical environments. It still
remains a significant challenge to understand how rotation
and concomitant magnetic fields will lead to jets in the
context of core collapse. 

Many groups are beginning to do rotating, magnetic simulations, 
but it is very challenging numerically to achieve the kind
of numerical resolution needed to resolve the MRI (Obergaulinger 
et al. 2006 and references therein). Given that practical 
limitation, many groups have elected to do simulations with 
exaggerated initial magnetic fields and/or fields with 
exaggerated poloidal structure.  Poloidal fields coupled with shear 
will naturally amplify the field strength by field-line wrapping. 
This is a generically slow process since it depends linearly on 
the winding.  To get ``useful" results it is necessary to have 
the poloidal configuration be significantly strong. Neither
a weak poloidal configuration nor a strong toroidal 
initial configuration are likely to produce interesting
jet-like behavior in the absence of a numerical capability
to follow the growth of the non-axisymmetric MRI that
can and will grow from a modest initial toroidal magnetic
field.  Calculations assuming an initially strong poloidal
field can produce spectacular numerical results (for a recent 
example, see Burrows et al. 2007), but it is not clear that these 
simulations reflect what Nature is doing. The initial field is
likely to be modest and substantially toroidal, not poloidal, since 
it is likely to have been generated by shear in the progenitor star. 
The same is true, of course, for the field generated by the MRI or
other dynamo processes in the proto-neutron star. The challenge
is to understand whether and how a jet is formed from an 
initially modest toroidal magnetic field. Note again that the
initial iron core rotation might be modest. If SASI-like processes
spin up the neutron star independent of the initial core rotation
and the rotation is enhanced during de-leptonization, there is 
still plenty of time and opportunity to affect the explosion 
dynamics with MHD jet and magnetoacoustic phenomena.

\section{Conclusions}

Core collapse supernovae, including SN 1987A, show departures from 
both spherical and axial symmetry. There is substantial evidence
that the explosion of SN 1987A itself was truly bi-polar, but with
important and interesting departures from axisymmetry. 
The proto-neutron star that forms after core collapse will be rotating 
and will have a concomitantly strong, predominantly toroidal magnetic 
field. These conditions might form a jet, but they are also conditions 
that could  give rise to a strong, magnetoacoustic wave-generating 
engine. Supernovae make a loud noise!

The role of non-axisymmetric instabilities may help to explain 
the spectropolarimetry of core collapse supernovae and may also 
have major implications for supernova physics as well as for
pulsar magnetic fields and spins.


\begin{theacknowledgments}

JCW and JRM are exceedingly grateful for the fruitful
collaboration with the members of our VLT spectropolarimetry
team, Dietrich Baade, Lifan Wang, Peter H\"oflich, 
Fernando Patat and Alejandro Clocchiatti and to the
wonderful facilities and crew at the VLT who make the
observations possible. JCW and JRM are supported in
part by NSF Grant AST--0406740.

\end{theacknowledgments}



\bibliographystyle{aipproc}   





\end{document}